\documentclass[conference]{IEEEtran}
\usepackage{xcolor}
\usepackage{graphicx}
\usepackage{cite}

\usepackage[bookmarks=false]{hyperref}
\ifCLASSINFOpdf
\else
\fi
\begin{document}
%

\title{Storing and retrieving wavefronts with resistive temporal memory}



\author{\IEEEauthorblockN{Advait Madhavan\IEEEauthorrefmark{1},
Mark D. Stiles\IEEEauthorrefmark{2}}
\IEEEauthorblockA{Physical Measurement Laboratory, National Institute of Standards and Technology\IEEEauthorrefmark{1}\IEEEauthorrefmark{2}\\
Institute for Research in Electronics and Applied Physics, University of Maryland, College Park\IEEEauthorrefmark{1}\\
Email: \IEEEauthorrefmark{1}advait.madhavan@nist.gov,
\IEEEauthorrefmark{2}mark.stiles@nist.gov}}


%


\maketitle

\begin{abstract}



We extend the reach of temporal computing schemes by developing a memory for multi-channel temporal patterns or ``wavefronts." This temporal memory re-purposes conventional one-transistor-one-resistor (1T1R) memristor crossbars for use in an arrival-time coded, single-event-per-wire temporal computing environment. The memristor resistances and the associated circuit capacitances provide the necessary time constants, enabling the memory array to store and retrieve wavefronts. The retrieval operation of such a memory is naturally in the temporal domain and the resulting wavefronts can be used to trigger time-domain computations. While recording the wavefronts can be done using standard digital techniques, that approach has substantial translation costs between temporal and digital domains. To avoid these costs, we propose a spike timing dependent plasticity (STDP) inspired wavefront recording scheme to capture incoming wavefronts. We simulate these designs with experimentally validated memristor models and analyze the effects of memristor non-idealities on the operation of such a memory. 

\end{abstract}


%
\IEEEpeerreviewmaketitle

\section{The need for Temporal Memory}

The three pillars that form the foundation of any computing system are computation (for processing), input/output (I/O) (for sensing and feedback), and memory (for storage). In the context of single-spike-per-wire, arrival-time coded computation, circuits that allow sensing and processing natively in the temporal domain are already being researched. Dynamic vision sensor (DVS) cameras \cite{lichtsteiner2008128}, time-to-first-spike (TTFS) vision sensors \cite{lenero2010signed}, and address event representation (AER) ears \cite{chan2007aer} are a few examples of the sensing systems that natively encode information in the temporal domain. On the computational side, a space-time computing approach \cite{smith2018space} has been proposed as a novel paradigm that encodes information in the relative arrival time between input events. A direct implementation of such a paradigm with off the shelf complementary-metal-oxide-semiconductor(CMOS) technology has also been proposed and demonstrated \cite{madhavan2014race,madhavan20174}. The active research on two of the three pillars increases the urgency to develop  of a memory that natively operates in the time domain. As of now, no such memory exists. 

\begin{figure}
    \centering
    \includegraphics[width=0.9\linewidth]{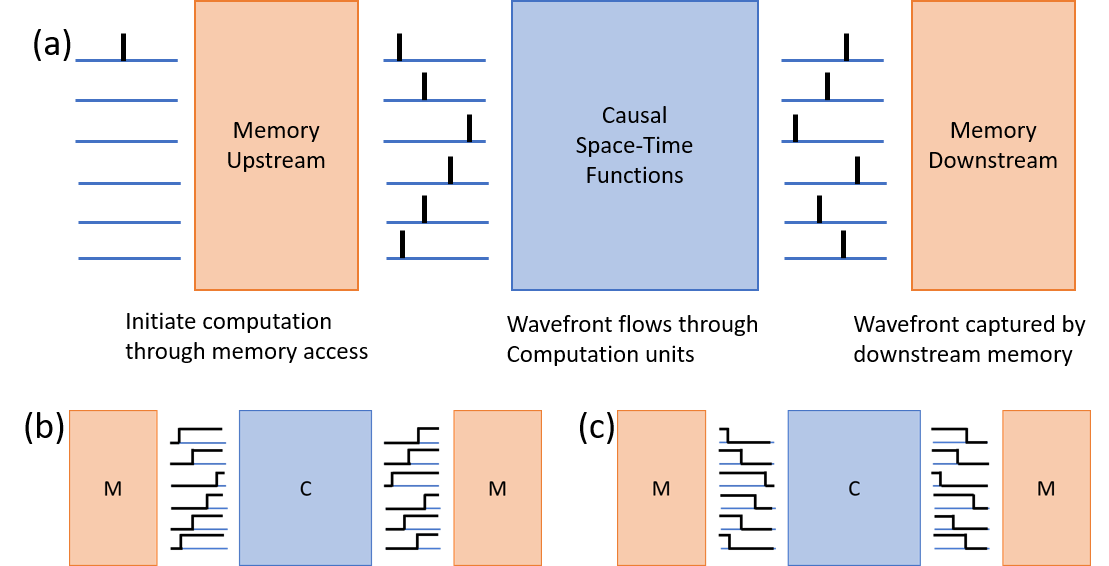}
    \caption{A temporal computation procedure: Panel (a) shows an upstream memory through which computation is initiated by triggering an event at the input at a given time. The memory outputs a sequence of events on different lines which encodes the temporal wavefront that is stored in the first memory location. The output wavefront from the computational unit is read in by the downstream memory and stored in a memory location of choice. Panels (b) and (c) depict race-logic-like implementations that represent events with digital rising or falling edges instead of spikes. }
    \label{fig:temp_comp}
\end{figure}

The biological motivation behind single-spike-per-wire temporal computation can be traced back to Thorpe and Imbert's work \cite{thorpe1989biological} arguing that the speed of processing of the visual system is too fast for a rate-coded interpretation of neural computation to be feasible \cite{thorpe1996speed}. Instead, they proposed a wavefront based computing approach that encodes information in the relative arrival time between a volley of spikes \cite{thorpe2001spike} as is shown in Fig.~\ref{fig:temp_comp}(a). This information representation is radically different from the conventional Boolean one and opens up a vastly different trade-off space of possible computing architectures\cite{madhavan2014race,madhavan20174,RaceTrees, smith2018space, najafi2018low,lagorce2016hots}. 

The temporal domain allows for different encoding schemes, among which two have been well studied \cite{thorpe1990spike,vanrullen2005spike,thorpe1998rank}. One is a timing code, where the exact delays between the spikes carry information \cite{rullen2001rate, vanrullen2005spike}. A more relaxed version of such an approach, though less dense in coding space, is a rank order code, in which only the relative orderings of the spikes carry information \cite{thorpe1998rank,rullen2001rate}. Though sparser than their exact timing counterparts, order codes have been shown to contain appreciable information capacity while still maintaining robustness to variations in individual spike timing \cite{thorpe1998rank}. Though the first temporal networks were studied as early as 1990 \cite{thorpe1990spike}, more recently, rank order codes have been studied in the context of deep spiking neural networks \cite{tavanaei2018deep,kheradpisheh2018stdp,mozafari2019bio}. Such single-event-per-wire based (also known as non-leaky integrate and fire) models have been trained with modified STDP algorithms. Early results \cite{diehl2015fast, kheradpisheh2018stdp} report comparable accuracies to deep learning networks but with  small network sizes.

Building hardware implementations of such systems requires physical realizations of events. Race logic \cite{madhavan2014race} is a temporally coded logic family that takes advantage of the simplicity of the digital domain and represents events with rising or falling digital edges (as shown in Figs.~\ref{fig:temp_comp}(b,c)), instead of spikes. Fig.~\ref{fig:temp_comp}(b,c) show rising and falling edge versions of architectures of such a temporal computer, built using the race logic encoding scheme.  Computations can be initiated by memory access to the upstream memory, which recalls a stored temporal wavefront. This wavefront flows through computational units that implement arbitrary causal functions such as the ones described in \cite{RaceTrees, smith2018space, najafi2018low}. Lastly the downstream memory gets triggered with the first arriving edge and captures the incoming wavefront.

\begin{figure}
    \centering
    \includegraphics[width=0.9\linewidth]{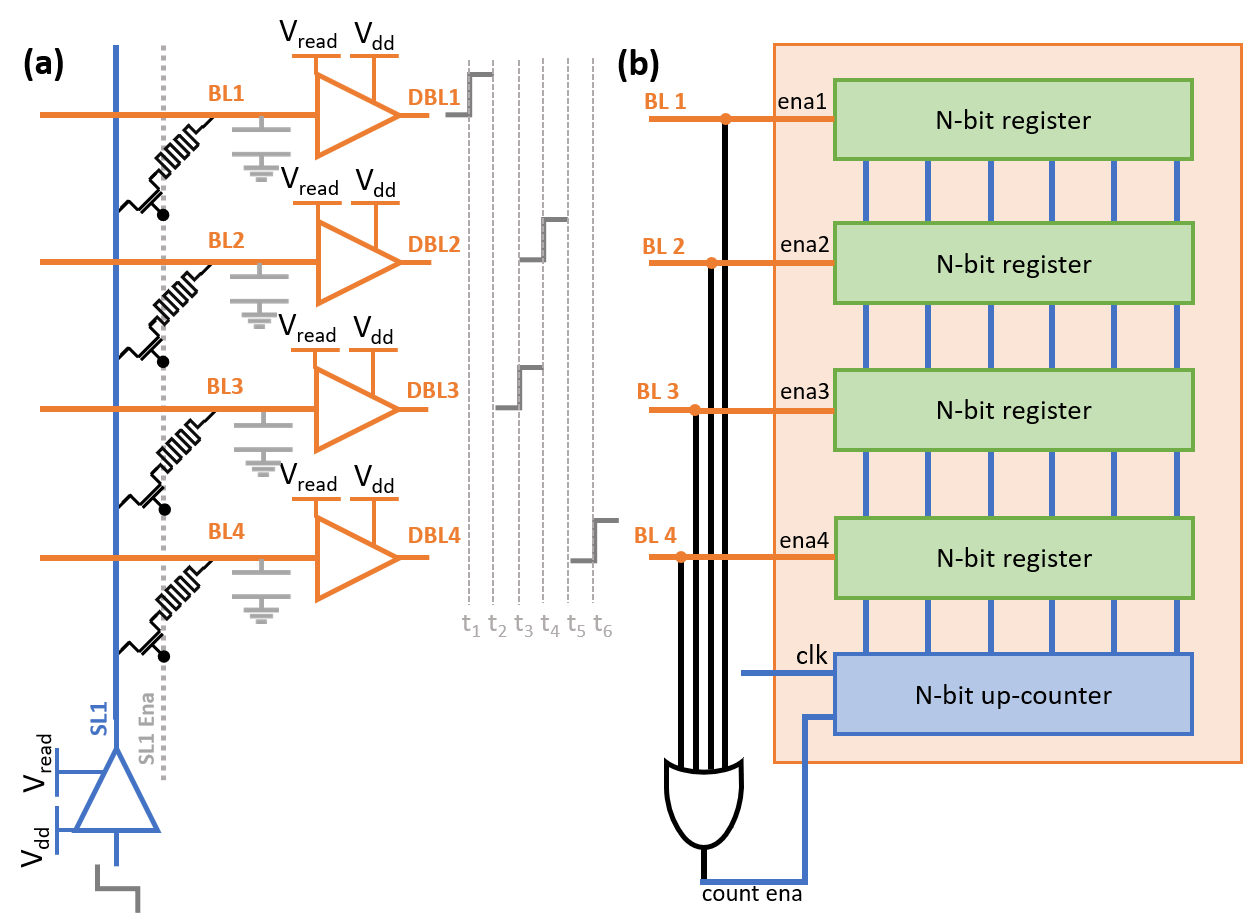}
    \caption{A simple temporal memory circuit: Panel (a) shows a read/recall operation where a rising edge is presented at the input of the source line driver (level shifter). This results in different charging rates of the various bit lines, determined by the respective cross-point devices, resulting in an output wavefront. This wavefront enters the bitlines of panel (b) where the OR gate detects the first arriving edge, starting the up-counter. The incoming rising edges on the bit-lines  latch the counted values and hence store the incoming wavefront.}
    \label{fig:simple_circuit}
\end{figure}

In this paper we present the design of a memory that fits seamlessly into a temporal computation procedure as described in figure \ref{fig:temp_comp}. We do this by performing a translation between static memory and timing signals, through tunable memristor RC time constants.  Section II describes how 1T1R memristive crossbars can be used to create wavefronts that have been stored in them. We describe how such an approach can interrogate the memristive state with more energy efficiency than conventional techniques. Section III describes how the relative timing information in a wavefront can be captured through standard digital circuit techniques which then invokes specific write circuitry to tune the memristor resistances to the corresponding captured digital values. This domain shifting, from analog to digital and back, has significant overhead associated with it. We then describe a proposed solution to natively capture wavefronts directly into memristors.

\section{Recalling wavefronts stored in a memristor crossbar}


An ideal temporal memory would be one that could be directly interfaced with digital components in a temporally coded environment where rising edges are used to demarcate events. Figure \ref{fig:simple_circuit}(a) shows a single column of such a memory, which uses a 1T1R memristor crossbar as its fundamental component. Each row behaves as an output bit line and each column behaves as the input source line. When a rising edge arrives through an enabled source line, it charges the output bit line (BL) MOS capacitor (shown in Fig.~\ref{fig:simple_circuit}), through the memristor, until a threshold is reached, causing a rising edge at the digital bit line (DBL). Using such a circuit, the values of the memristive states can be directly read out at as a wavefront of digital rising edges, also known as wavefront recalling. This is shown in Fig.~\ref{fig:simple_circuit}, where a linear variation in memristive values leads to a linearly spaced output wavefront. 

\begin{figure*}
    \centering
    \includegraphics[width=0.9\linewidth]{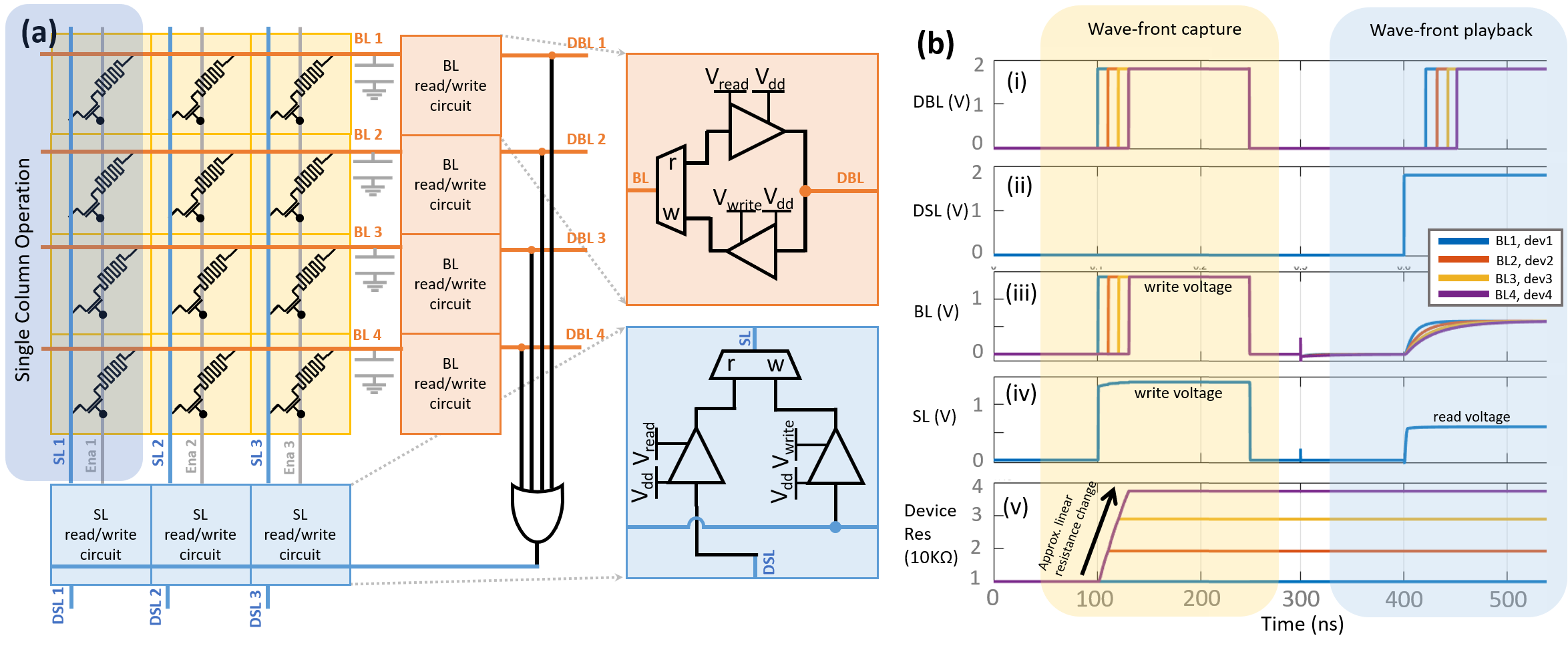}
    \caption{Simulation results for recalling and recording wavefronts with resistive temporal memory: Panel~(a) shows a 3X4 resistive temporal memory with the yellow cells representing the conventional 1T1R array, while the blue cells and orange cells represent the source and bit line augmentations that allow using such an array in a temporal context. Note that the level shifters shown in the zoomed in cells are digital, with the read out cells being tri-state, while the multiplexers that explicitly depict the read and write circuit paths are pass gate based. Panel~(b) shows the SPICE simulation results of recording a wavefront and subsequently recalling it for the single column in Panel~(a) that is highlighted in blue. The wavefronts are superimposed over each other to save space. In the capture phase, the bit-lines (BLs) are used as inputs, while the source lines(SLs) are controlled by the OR gate through the write path. In the recall phase, the source lines are used as the inputs with the bit lines being the outputs. Panels~(i) and (ii) show the digital signal values, while panels~(iii) and (iv) show the internal BL and SL values. Panel~(v) shows the device state change during the capture phase. }
    \label{fig:arch_results}
\end{figure*}

Though the structure of the crossbar remains the same, the way it is used in this work differs in some important ways from conventional approaches. When used in a multilevel memory or feed-forward inference context, a static read voltage is applied across the device (pinned with a sense/measurement amplifier) while the resultant current is summed and measured. Hence, the energy efficiency in these approaches improves the larger the $R_{\rm on}$ and $R_{\rm off}$ resistances become. In contrast, in this RC charging based recall mode of operation, the voltage drop across the device is not static, because the voltage on the output capacitor changes during a read operation (Fig.~\ref{fig:arch_results}(b)(iii)). 

This changing voltage has two advantages. First, and more important, it decouples the energy cost per read operation from the value stored in the memristor. Independent of the state of the device, a single read operation consumes $CV_{\rm read}^2 (\approx 600~{\rm fJ})$ of energy per line, with $CV_{\rm read}^2/2$ lost due to joule heating across the memristor and $CV_{\rm read}^2/2$  stored on the capacitor. This data independent energy cost allows memristors to be used in the high conductance regime, without incurring the increased energy cost. Circuit and architectural designs can then take advantage of the high conductance regime, where the behaviour of the device is more linear, repeatable and less susceptible to variation. Recently, for very low resistance states, the device to device variation has shown to be $\leq 1\%$ \cite{li2018analogue}. The second advantage is that the degree of read disturb on the device is reduced as the full read voltage is applied across the device for a relatively short period of time. 

To enable easy interface with digital signal levels, level-shifters are required to translate between the memristor read voltages ($V_{\rm read}$)  and digital voltage levels ($V_{\rm dd}$). This shifting down process can be implemented with regular inverters but the shifting up process requires either current mirror based level-shifters or cross coupled level-shifters. The current mirror based designs have a smoother response, and consume static power while the cross coupled versions are more power efficient, but have a more complicated response. 

\begin{figure}
    \centering
    \includegraphics[width=0.6\linewidth]{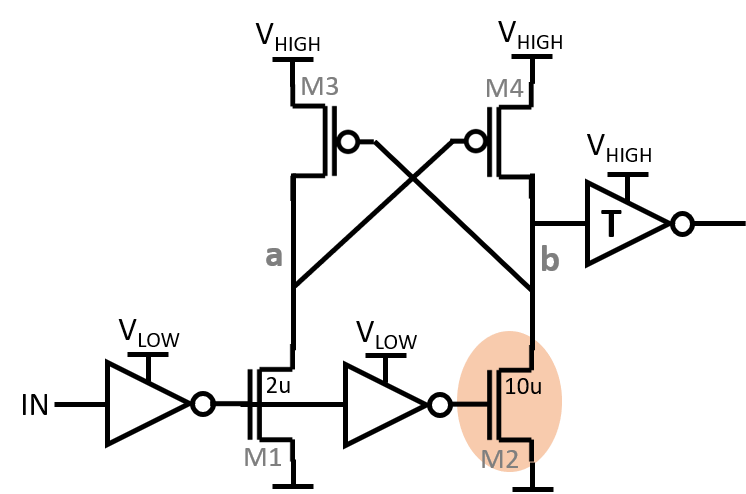}
    \caption{Asymmetric rising edge level shifter: Here transistor M2 is sized larger than its counterpart M1 such that node ``b'' is pulled down faster with little competition from M1 via node ``a''. The inverter with a ``T'' inside represents a tri-state buffer. }
    \label{fig:asym_lvl_shift}
\end{figure}

The cross coupled topology is representative of a positive feedback loop between transistors M1-M4 (Fig.~\ref{fig:asym_lvl_shift}). This positive feedback loop itself has a time constant that varies with the current charging the input node. This variable time constant can add timing uncertainties  that are data dependent and could cause errors. One way to avoid this problem is to take advantage of the one sided nature of this information encoding. Using rising edges only determines the transistor that is responsible for the pull-down so it can be sized accordingly larger. This approach makes the response of the level shifter more uniform. 

\section{Capturing Wavefronts: Digital vs Native}


A functionally correct digital timing measurement approach to record wavefronts is shown in Fig.~\ref{fig:simple_circuit}(b). High speed up-counters can be used for time scales on the order of 1~ns to 50~ns, while vernier delay lines, which extend the precision to the order of a single-inverter-delay, can be used for more precise measurements \cite{dudek2000high}. Using race logic principles, the first arriving edge is detected with an OR gate, which signals the beginning of the timing measurement system (counter or vernier delay line). With each subsequently arriving rising edge, the corresponding count value is captured in a register bank. An AND gate signals the last arriving input, at which time the recording process comes to an end with a digital representation of the relative arrival times with stored in a temporary register bank. These values can be used as targets for a closed loop feedback programming approach \cite{alibart2012high} that writes the corresponding values into the correct memory column.

To increase the energy efficiency of wavefront recording we eliminate the need to translate between encoding domains by using the ability to change memristor resistances with applied voltage pulses. This native approach to capturing wavefronts, results in a  more natural and energy efficient, albeit more error prone implementation. In a time coded information representation, in which the plastic  memristor resistances explicitly encode tunable delays, STDP-like behaviour can be used to record wavefronts as shown in Fig.~\ref{fig:arch_results}. In this approach, the first arriving edge is conceptually treated as the ``post'' edge. The circuit then applies backward pulses of variable lengths across the memristors proportional to the \emph{difference in timing} between this first-arriving ``post'' event and the later-arriving events, which can be thought of as ``pre'' events. The device with the largest difference between ``pre'' and ``post'' events, has the maximum conductance change, and hence the highest resistance. When a wavefront is then recalled, the highest conductance device responds first and the most resistive one responds last, preserving the wavefront shape. 

Simulation results for such a procedure are shown in Fig.~\ref{fig:arch_results}(b). These simulations are performed in a 180~nm process node, with a 1.8~V power supply. The memristor models used are from \cite{chen2015compact}, and are modelled based on experimental measurements reported in \cite{chen2015compact, jiang2014verilog}.  
The wavefront recording operation proceeds by first initializing the column in question, (column 1, shown in figure \ref{fig:arch_results}(a)), with all memristors set to the ON state($\approx$ 10~k$\Omega$) and the enable line (Ena1) activated. This can be seen in the first 100~ns of figure \ref{fig:arch_results}(b)(v) with all devices having the same impedance. The write path through the multiplexers, as shown in figure \ref{fig:arch_results}(a), is also activated, such that the OR gate controls the source line (SL). 

The wavefront (having a dynamic range of 40~ns) to be recorded is presented at the digital bit lines (DBLs), which behave like the input in this phase of operation. Similarly to the digital case, the first arriving rising edge is detected by an OR gate, which triggers the application of an appropriate write voltage ($V_{\rm write}\approx$ 1.4~V), through the multiplexer, to the source line (SL). The bit-lines (BLs) of the array are operated in the write voltage regime with rising edges level shifted down from $V_{\rm dd}$ to $V_{\rm write}$. Each device sees the difference in voltage between the source line(figure~\ref{fig:arch_results}(b)(iv)) and corresponding bit lines(figure~\ref{fig:arch_results}(b)(iii)) applied across it. For the device corresponding to the first arriving edge, both its source line and bit line go high at the same time, so there is no change in the memristive state. Meanwhile, the other devices experience a reverse $V_{\rm write}$ voltage across them, since their edges haven't arrived yet. They experience this voltage for the difference in time between their corresponding edge and the first arriving edge, hence causing a change in the memristive state proportional to the relative times between the inputs. 

Once appropriate pulse lengths have been successfully applied across the devices, a copy of the input wavefront should be captured into the memristive device states. The last arriving edge signals the end of the recording operation and the circuit is reset with an external reset pulse. The reset pulse (not shown in the figures), discharges the crossbar without affecting the device state. The array is now ready for playback.

\section{Discussion}

While such a wavefront recording approach seems feasible, some problems arise in the context of exact timing codes. First, the relationship between memristor resistance and relative output timings for recalling the wavefront is linear, arising directly from the $t \propto RC$ relationship. On the other hand, for recording the wavefront, the relationship between memristor conductance and voltage pulse duration is not linear, and depends on material properties. Since most memristive state change dynamics are governed by transitions over energy barriers, the effectiveness of a fixed voltage to change the device state drops logarithmically. In the wavefront recording process, a linearly spaced input wavefront will end up creating a logarithmically spaced resistive change, which when recalled would create a logarithmically spaced output wavefront. This problem is fundamental, being governed by the exponential nature of Boltzmann statistics and energy barriers. 

In order to get linear behavior out of such a device, it must operate in a regime where the Taylor series expansion of its behavior has small higher order coefficients, so that it can be approximated as linear. Such behavior can be seen for a range of voltages where a sharp pulse ($\leq$ 40~ns) across the device creates a linear change in the devices state (from 10~k$\Omega$ to 40~k$\Omega$), which tapers off if the pulse is applied for a longer time. Here, as shown in figure \ref{fig:arch_results}(e)(v), our pulse duration is calibrated to access that specific linear region of the memristor IV characteristics, and therefore does not access the complete available dynamic range. 

The reduced range is not detrimental and depends on the quality of the memristive devices being used. Multiple groups have shown 5 bit or more resolution in limited resistance ranges with the low resistance state incurring, programming cycle to cycle variations as low as $4.2~\%$ and device to device variation as low as $4.5~\%$ \cite{cai2019fully}. For very low resistances (between 1~k$\Omega$ and 10~k$\Omega$), even lower variation numbers have been reported ($\leq 1~\%$ \cite{li2018analogue}). Such technological improvements allow us to extract 4 to 5 bits of precision, even from a reduced dynamic range.

A second difficulty for exact timing codes is that the time scales of recording and of recalling need to match. For example, the resistance change created by 10~ns pulses in the recording process, should create 10~ns spaced edges when recalled. While the former is a material property and cannot be changed by circuit techniques, the latter can be addressed by adding a digitally programmable capacitances ($\approx$ 1~pF, in the current simulation) on the output line to correctly scale the timescale. For small array sizes such a capacitance can take up extra area, but as the array is scaled to more representative sizes, the crossbar, transistor-drain and driver capacitances will contribute significantly to this capacitance. Array scaling will also require scaling of the array drive circuits, especially with the high conductance regime operation. Though the write drivers on the bit line do not need to be adjusted, but the source line write driver will have to be designed to support $N$ memristors in parallel during the capture phase. Future work includes a more detailed scaling analysis accounting for crossbar wire resistances and capacitances. 

An important point to note is that rank order codes are more tolerant to the aforementioned concerns than exact timing codes. Logarithmic compression preserves order, and variable capacitances can be used with order codes to stretch the logarithmically compressed stored values. This allows enough write pulse duration to still change state on the next write operation. This makes rank order codes a more robust and error tolerant encoding for this kind of a temporal memory.

\section{Conclusion}



In this work we have proposed and validated through simulation, a single-event-per-wire temporal memory that operates in the sub 50~ns timing range while utilizing the low variability, low resistance states (10~k$\Omega$ to 40~k$\Omega$) of memristive devices. We show how our recalling/playback operation has an energy cost of about 600~fJ per line, whose magnitude is independent of the device conductance. Rank order coded architectures seem to be the more promising encoding schemes for such an approach due to their error tolerance. Though many challenges remain, we believe that this is a first step towards realizing temporal memories that can work synergistically with tomorrow's temporally coded computing architectures.





\bibliographystyle{IEEEtran}
\bibliography{refs}


%

\end{document}